\documentclass[a4paper,11pt]{article}
\pdfoutput=1 % if your are submitting a pdflatex (i.e. if you have
\usepackage{jcappub} % for details on the use of the package, please
\usepackage{bm}

\usepackage[T1]{fontenc} % if needed

\begin{document}

\title{\boldmath Primordial black hole formation from collapsing domain walls with full general relativity}

%% %simple case: 2 authors, same institution
%% \author{A. Uthor}
%% \author{and A. Nother Author}
%% \affiliation{Institution,\\Address, Country}

% more complex case: 4 authors, 3 institutions, 2 footnotes
\author[a,b]{Naoya Kitajima}
%\author[c]{S. Econd,}
%\author[a,2]{T. Hird\note{Also at Some University.}}
%\author[a,2]{and Fourth}

% The "\note" macro will give a warning: "Ignoring empty anchor..."
% you can safely ignore it.

\affiliation[a]{Frontier Research Institute for Interdisciplinary Sciences, Tohoku University, \\
6-3 Azaaoba, Aramaki, Aoba-ku, Sendai 980-8578, Japan}
\affiliation[b]{Department of Physics, Tohoku University, \\
6-3 Azaaoba, Aramaki, Aoba-ku, Sendai 980-8578, Japan}

% e-mail addresses: one for each author, in the same order as the authors
\emailAdd{naoya.kitajima.c2@tohoku.ac.jp}

\abstract{
We study the dynamics of isolated closed domain walls with 3+1 numerical relativity. A closed wall shrinks due to its own surface tension, and its surface energy is converted to the kinetic energy, leading to implosion. Then, it can result in the formation of a black hole. First, we focus on spherically symmetric closed domain walls and clarify whether they finally evolve into black holes. Naively, the wall can collapse if its thickness is smaller than the Schwarzschild radius which is determined by the initial surface energy. Our numerical results support this naive criterion for the black hole formation, and indicate that more than 80\% of the initial wall energy falls into the black hole. We also investigate the nonspherical collapse by considering the ellipsoidal configurations for the closed domain walls, and it turns out that black holes can be formed even when the ratio of semi-major to semi-minor axes is 1.5.
}

%\begin{flushright}
%TU-XXXX
%\end{flushright}

\maketitle
\flushbottom

\section{Introduction} \label{sec:intro}

The domain wall is a sheet-like topological defect arising from the spontaneous breaking of discrete symmetry \cite{Zeldovich:1974uw,Kibble:1976sj}. When such a symmetry is broken in the early universe, domain walls form a network, and it evolves following the so-called scaling law in the expanding universe. The dynamics of each domain wall is fully nonlinear, but once the system enters the scaling regime, the network shows a simple self-similar evolution, that is, roughly one domain wall intersects one Hubble volume \cite{Press:1989yh,Hindmarsh:1996xv,Garagounis:2002kt,Oliveira:2004he,Avelino:2005kn,Leite:2011sc}. Consequently, the energy density of the wall network decreases as $\rho_{\rm DW} \sim \sigma_{\rm DW} H$, where $\sigma_{\rm DW}$ is the wall tension and $H$ is the Hubble parameter. In general, the tension is a constant determined by fundamental parameters, and thus $\rho_{\rm DW} \propto 1/t$ in the universe dominated by radiation or matter. Therefore, the energy density of the wall decreases slower than the background energy density, implying that the domain wall eventually dominates the universe unless the energy scale of the wall is sufficiently low. It spoils the success of standard cosmological scenario, and thus it is called the cosmological domain wall problem.

The domain wall problem can be evaded if the wall network is annihilated away before the domination. The domain wall annihilation can be realized by introducing the bias \cite{Gelmini:1988sf,Larsson:1996sp}. For example, if the minima of the scalar potential are not exactly degenerate and there is a small difference between the height of these potential minima, which we denote $\Delta V$, there is a pressure to decrease the volume of the false vacuum region. When the pressure becomes comparable to the wall energy, the wall network decays within one Hubble time. This is one of the solutions of the domain wall problem.

The domain wall has high energy concentration in its core and induces strong gravity. In fact, the dynamics of the wall network continuously emits gravitational waves through the anisotropic stress of the scalar field \cite{Gleiser:1998na}. Since the energy density fraction of the domain wall gradually increases compared with the background value, the emission of gravitational waves is most efficient when the wall network decays. Thus, the resultant spectrum of gravitational waves has a characteristic peak which can be probed by gravitational wave observations \cite{Hiramatsu:2010yz,Kawasaki:2011vv,Hiramatsu:2013qaa,Kitajima:2015nla,Higaki:2016jjh,Nakayama:2016gxi,Kitajima:2023cek,Ferreira:2024eru,Dankovsky:2024zvs}.

In this paper, we focus on the primordial black hole (PBH) \cite{Zeldovich,Hawking:1971ei,Carr:1974nx} as another gravitational remnant from domain walls. The formation of PBH has been extensively studied, as it can be an imprint of the early universe, and indeed observations can put stringent constraints on the abundance of PBH. See \cite{Carr:2009jm,Carr:2016drx,Sasaki:2018dmp,Escriva:2022duf,Byrnes:2025tji} for reviews. The PBH formation from the domain wall has been studied in the literature \cite{Ipser:1983db,Widrow:1989fe,Widrow:1989vj,Rubin:2000dq,Tanahashi:2014sma,Garriga:2015fdk,Deng:2016vzb,Ge:2019ihf,Liu:2019lul,Eroshenko:2021sez,Ge:2023rrq,Dunsky:2024zdo}.\footnote{The formation of PBH from the string-wall system in the axion model is studied in \cite{Vachaspati:2017hjw,Ferrer:2018uiu,Gelmini:2022nim,Gelmini:2023ngs}. In addition, the PBH formation from density fluctuations sourced by the dynamics of the wall network is pointed out in \cite{Lu:2024ngi,Lu:2024szr}.
}
In particular, the isolated closed domain wall is considered as a seed of PBH. Once such an object is formed, it starts to shrink due to the surface tension. Then, the surface area of the wall decreases continuously and the surface energy is converted to the kinetic energy, which finally leads to implosion. Suppose that the minimum possible size of the system is smaller than the Schwarzschild radius, a black hole can be formed.

To follow the dynamics of the black hole formation, the fully nonlinear nature of gravity should be taken into account. In this paper, we numerically study the dynamical process of the black hole formation from closed domain walls in a fully general relativistic approach. Namely, we employ the 3+1 formulation of numerical relativity. This approach has been applied to the PBH formation from primordial density (curvature) fluctuations \cite{Yoo:2020lmg,deJong:2021bbo,Escriva:2021aeh,Yoo:2021fxs,deJong:2023gsx,Yoo:2024lhp,Escriva:2024lmm}. See also \cite{Aurrekoetxea:2024ypv} for a review on cosmological applications of numerical relativity. General relativistic simulations of the PBH formation from domain walls are performed in \cite{Deng:2016vzb}. 
This previous work focuses on the situation in which closed domain walls are nucleated during inflation and all of them can collapse into black holes after the horizon crossing. In this paper, we focus on the situation in which such closed walls are rare objects, implicitly assuming the Kibble mechanism for the domain wall formation \cite{Kibble:1976sj}, and clarify the criterion for the PBH formation. 
The condition for the PBH formation in such a situation is obtained in \cite{Dunsky:2024zdo} based on semi-analytic consideration with flat space lattice simulations. We confirm the result of this previous study with fully general relativistic simulations.

The simplest configuration for the closed wall is a spherical shell. However, there is no physical process to make it spherical for the initial shape, and thus nonspherical configurations should be taken into account to correctly estimate the PBH formation probability. The nonspherical collapse for the PBH formation has been studied in \cite{Kuhnel:2016exn,Yoo:2020lmg,Yoo:2024lhp,Escriva:2024lmm} in the case of the PBH formation from primordial density fluctuations. In particular, even small deviations from sphericity can prevent the PBH formation \cite{Escriva:2024lmm}. The effect of the nonsphericity in the domain wall case is discussed in Ref. \cite{Dunsky:2024zdo} with flat space lattice simulations, showing that the efficiency of the PBH formation is reduced but not severe. In this paper, we also examine the PBH formation with ellipsoidal wall configurations using numerical relativity.

This paper is organized as follows. In Sec.~\ref{sec:domain_wall}, we introduce the domain wall model and show the naive criterion for the PBH formation from closed domain walls. Sec.~\ref{sec:setup} contains the setup for our numerical analysis, including the formulation of numerical relativity, initial conditions, numerical methods, and parameters. Sec.~\ref{sec:numerical} shows our numerical results in the case of both spherical and nonspherical collapses. There, we summarize the viable parameter space for the PBH formation. Sec.~\ref{sec:discussion} is devoted to the discussion.

\section{PBH formation from domain walls} \label{sec:domain_wall}

\subsection{Domain wall model}

We consider the model with a real scalar field, $\phi$, which constitutes the domain wall. Then, the action of the matter sector is given by
\begin{align} \label{eq:S}
S = \int d^4 x \sqrt{-g} \left[ -\frac{1}{2} \nabla^\mu \phi \nabla_\mu \phi - V(\phi) \right],
\end{align}
where $\nabla_\mu$ is the covariant derivative associated with the 4-dimensional metric $g_{\mu\nu}$, $g$ is the determinant of this metric, and $V(\phi)$ is the scalar potential. In this paper, we focus on the so-called $Z_2$ domain wall model with the following double-well potential,
\begin{align}
    V(\phi) = \frac{\lambda}{4}(\phi^2 - v^2)^2,
\end{align}
where $\lambda$ is the dimensionless self-coupling constant and $v$ is the vacuum expectation value of the scalar field. This potential has two degenerate vacua, $\phi = \pm v$, and thus if the $Z_2$ symmetry is spontaneously broken, the domain wall arises as a topological defect.

\begin{figure}[tbp]
\centering 
\includegraphics[width=8.5cm,clip]{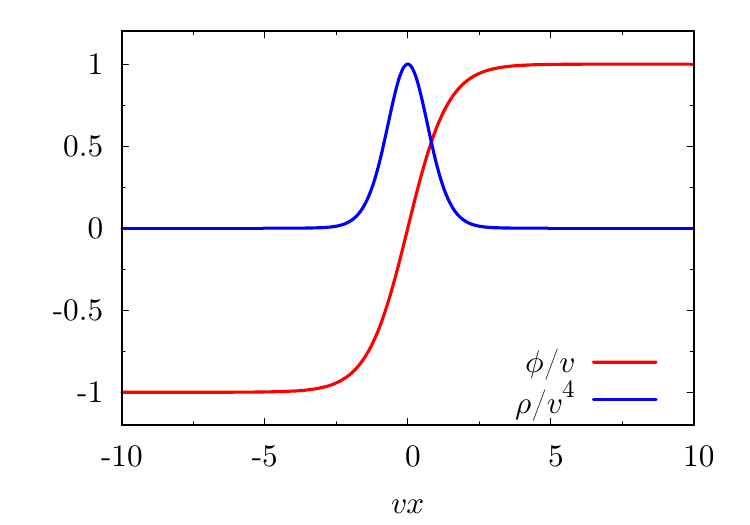}
\caption{\label{fig:scalarProfile} Spatial profiles of the scalar field (red) and the energy density (blue) for the $Z_2$ domain wall. We set $\lambda=1$.
}
\end{figure}

In the flat Minkowski spacetime, i.e. $g_{\mu\nu} = {\rm diag}(-1,1,1,1)$, the equation of motion for the scalar field is derived as follows,
\begin{align}
\ddot{\phi} - \nabla^2 \phi + \frac{\partial V}{\partial \phi} = 0,
\end{align}
where the overdot represents the time derivative.
For the 1+1 dimensional spacetime, or equivalently, if there is a shift symmetry along the remaining two axes (i.e. $\partial_y \phi = \partial_z \phi = 0$), there is a static solution given by
\begin{align}
\phi(x) = v \tanh \left( \sqrt{\frac{\lambda}{2}} v (x - x_0) \right),
\end{align}
with $x_0$ being some reference point. This solution also represents the static domain wall which is infinitely stretched in the $y$-$z$ plane. The energy density of the wall can be calculated through the scalar field energy density
\begin{align}
\rho = \frac{1}{2} \dot\phi^2 + \frac{1}{2} (\nabla\phi)^2 + V,
\end{align}
and the tension of the wall, defined by the energy per unit area, can be calculated by integrating the above energy density along the $x$-axis. Specifically, the tension and the thickness of the $Z_2$ domain wall are given by
\begin{align}
\sigma_{\rm DW} = \frac{4}{3} \sqrt{\frac{\lambda}{2}} v^3,~~\delta_{\rm DW} = \left(\sqrt{\frac{\lambda}{2}}v\right)^{-1}.
\end{align}
The profiles for the scalar field and the energy density are shown in Fig.~\ref{fig:scalarProfile}. 
The solution for the moving domain wall with a constant velocity $u$ can be obtained by taking the Lorentz transformation (boost) as follows
\begin{align} \label{eq:moving_dw}
    \phi(t,x) = v\tanh\left[ \sqrt{\frac{\lambda}{2}}v \gamma(x - u t - x_0) \right],
\end{align}
where $\gamma = (1-u^2)^{-1/2}$ is the Lorentz $\gamma$ factor. Note that when the wall velocity is ultra-relativistic in the observer's frame, i.e. $u \approx 1$ and $\gamma \gg 1$, the wall thickness suffers a significant Lorentz contraction and the wall tension increases as $\sigma_{\rm DW} \approx \gamma \sigma_{\rm DW}^0$ with $\sigma_{\rm DW}^0$ being that of the static wall.

\subsection{PBH formation criterion} \label{subsec:PBH}

Let us consider the closed domain wall which is decoupled from the global domain wall network. Here, we consider the spherically symmetric closed wall as the simplest example of such a system. In a spherically symmetric system, the equation of motion for the scalar field in a flat Minkowski spacetime can be written as
\begin{align} \label{eq:eom_sph}
\ddot{\phi} - \frac{\partial^2 \phi}{\partial r^2} - \frac{2}{r} \frac{\partial \phi}{\partial r} + \frac{\partial V}{\partial \phi} = 0.
\end{align}
As long as the third term in the left hand side is negligible, the equation of motion is equivalent to that in the 1+1 dimensional spacetime, and the solution is given by Eq.~(\ref{eq:moving_dw}) with $x$ replaced by $r$.

Following Ref.~\cite{Dunsky:2024zdo}, here we derive the criterion for the PBH formation from the spherically closed domain wall.
It shrinks due to its own surface tension, and the initial energy stored in the wall is converted to the kinetic energy. 
Then, the wall is accelerated more and more as the surface area reduces, and finally it is maximally compressed to the size of the wall thickness as a result of implosion.
One can naively expect that if it enters the Schwarzschild radius, which is determined by the wall energy itself, a black hole is formed. 
Suppose that some fraction, $f$, of the initial wall energy contributes to the black hole mass, the Schwarzschild radius is given by $R_s = 2GM$ with $M = 4 \pi f R_0^2 \sigma_{\rm DW}$, which reads
\begin{align}
R_s = 2GM = \frac{4f}{3}\sqrt{\frac{\lambda}{2}} \frac{v^3 R_0^2}{M_{\rm pl}^2},
\end{align}
where $R_0$ is the initial radius of the closed wall.
A black hole is formed if $R_s$ is greater than the wall thickness $\delta_{\rm DW} \simeq (\sqrt{\lambda/2}v)^{-1}$.
Then, the naive criterion $R_s > \delta_{\rm DW}$ reads
\begin{align} \label{eq:PBHcriterion_naive}
\frac{R_0}{\delta_{\rm DW}} > \sqrt{\frac{3}{4f}}\frac{M_{\rm pl}}{v}.
\end{align}
However, as the wall is continuously accelerated due to the surface tension, the wall velocity becomes highly relativistic and the wall thickness suffers a Lorentz contraction.
As mentioned above, since the surface tension scales as $\sigma_{\rm DW} \propto \gamma$ for relativistic walls, the energy conservation, $R^2 \sigma_{\rm DW} = {\rm const}$, leads to $\gamma = (R_0/R)^2$ \cite{Widrow:1989vj}.
On the other hand, the plane wall solution (\ref{eq:moving_dw}) is no longer valid when the curvature term (the third term) in Eq.~(\ref{eq:eom_sph}) becomes comparable to other terms. It occurs when
\begin{align}
\Bigg| \frac{(2/r)\partial_r \phi}{\partial V/\partial \phi}\Bigg| \sim \frac{\gamma \delta_{\rm DW}}{R} \sim \frac{\delta_{\rm DW}}{R_0} \left(\frac{R_0}{R} \right)^3 \sim 1,
\end{align}
and thus, the plane wall description cannot be applied when the shell radius becomes smaller than the critical value: $R_* = (\delta_{\rm DW}/R_0)^{1/3}R_0$. 
The Lorentz factor at that time is then $\gamma_* = (R_0/\delta_{\rm DW})^{2/3}$ which determines the minimum possible wall thickness, $\delta_* = \delta_{\rm DW}/\gamma_* = \delta_{\rm DW} (\delta_{\rm DW}/R_0)^{2/3}$ \cite{Dunsky:2024zdo}.
Therefore, the above naive criterion is modified as follows,
\begin{align} \label{eq:PBHcriterion}
    \frac{R_0}{\delta_{\rm DW}} > \left( \frac{3}{4f} \right)^{3/8} \left(\frac{M_{\rm pl}}{v} \right)^{3/4}.
\end{align}
The collapse fraction $f$ can be determined by numerical simulations.

\section{Numerical setup} \label{sec:setup}

\subsection{3+1 formalism}

To take into account a fully general relativistic nature of the domain wall dynamics, we have to follow the local evolution of the metric. 
Here we employ the Arnowitt-Deser-Misner (ADM) or 3+1 formalism of general relativity \cite{Arnowitt:1959ah,Arnowitt:1962hi}, in which the spacetime metric is formally expressed as
\begin{align}
ds^2 = -\alpha^2 dt^2 +\gamma_{ij} (\beta^i dt+dx^i)(\beta^j dt + dx^j),
\end{align}
where $\alpha$ is the lapse, $\beta_i$ is the shift vector, $\gamma_{ij}$ is the induced metric on the spatial hypersurface $\Sigma_t$.
The lapse and the shift vector can be specified by the gauge fixing condition.
Since the Einstein equation is a set of second-order differential equations in time, we also need the time derivative of the spatial metric to solve the equation as an initial value problem. 
Then, the extrinsic curvature is introduced as 
\begin{align}
K_{\mu\nu} = -\gamma^\rho_\mu \gamma^\sigma_\nu \nabla_\rho n_\sigma,
\end{align}
where $\gamma^\mu_\nu$ is the projection tensor onto the spatial hypersurface $\Sigma_t$, and $n_\mu$ is a time-like vector perpendicular to $\Sigma_t$ given by
\begin{align}
n_\mu = (-\alpha,0,0,0) ~~~\text{and}~~~n^\mu = (\alpha^{-1},-\alpha^{-1}\beta^i).
\end{align}
Note that the extrinsic curvature is purely spatial and we denote it by $K_{ij}$ in what follows.
Practically, the extrinsic curvature is decomposed into the trace part ($K$) and the traceless part ($A_{ij}$) in the following way,
\begin{align}
K_{ij} = A_{ij} + \frac{1}{3}\gamma_{ij} K.
\end{align}
Moreover, we denote the spatial metric as the product of the conformal factor $\chi$ and the conformal metric $\tilde\gamma_{ij}$ as follows,
\begin{align}
\gamma_{ij} = \chi^{-1} \tilde\gamma_{ij},\quad \det \tilde\gamma_{ij} = 1.
\end{align}
The Einstein equation can be translated to a system of first-order differential equations in time for a set of variables $(\chi,\tilde\gamma_{ij},K,A_{ij})$ with the Hamiltonian and momentum constraints. In our simulations, we employ the so-called CCZ4 formulation with the moving puncture gauge.
The evolution equations together with the constraint equations and the gauge fixing condition are shown in the appendix \ref{sec:ccz4}.

The evolution equation for the scalar field should be rewritten in a general relativistic form.
Defining the conjugate momentum,
\begin{align}
\Pi = \frac{1}{\alpha}(\partial_t \phi-\beta^i \partial_i \phi),
\end{align}
the action (\ref{eq:S}) gives the following field equations in the 3+1 formalism,
\begin{align}
\partial_t \phi &= \alpha \Pi + \beta^i \partial_i \phi, \\[1mm]
\partial_t \Pi &= \beta^i \partial_i \Pi + \gamma^{ij} (\alpha \partial_i \partial_j \phi + \partial_j \phi \partial_i \alpha) + \alpha \left( K \Pi - \Gamma^k \partial_k \phi - \frac{\partial V}{\partial \phi} \right),
\end{align}
where $\Gamma^k = \gamma^{ij} \Gamma^k_{ij}$ with $\Gamma^k_{ij}$ being the Christoffel symbol with respect to the spatial metric $\gamma_{ij}$.
The energy momentum tensor of the scalar field is given by
\begin{align}
    T_{\mu\nu} = \nabla_\mu \phi \nabla_\nu \phi + g_{\mu\nu} \left( - \frac{1}{2} \nabla_\alpha \phi \nabla^\alpha \phi - V(\phi) \right),
\end{align}
and the energy density in the ADM formalism is defined as
\begin{align}
    \rho = n^\mu n^\nu T_{\mu\nu} = \frac{1}{2} \Pi^2 + \frac{1}{2} D_i \phi D_i \phi + V(\phi),
\end{align}
where $D_i$ denotes the covariant derivative associated with the spatial metric $\gamma_{ij}$.

\subsection{Initial condition}

\begin{figure}[tbp]
\centering 
\includegraphics[width=12cm,clip]{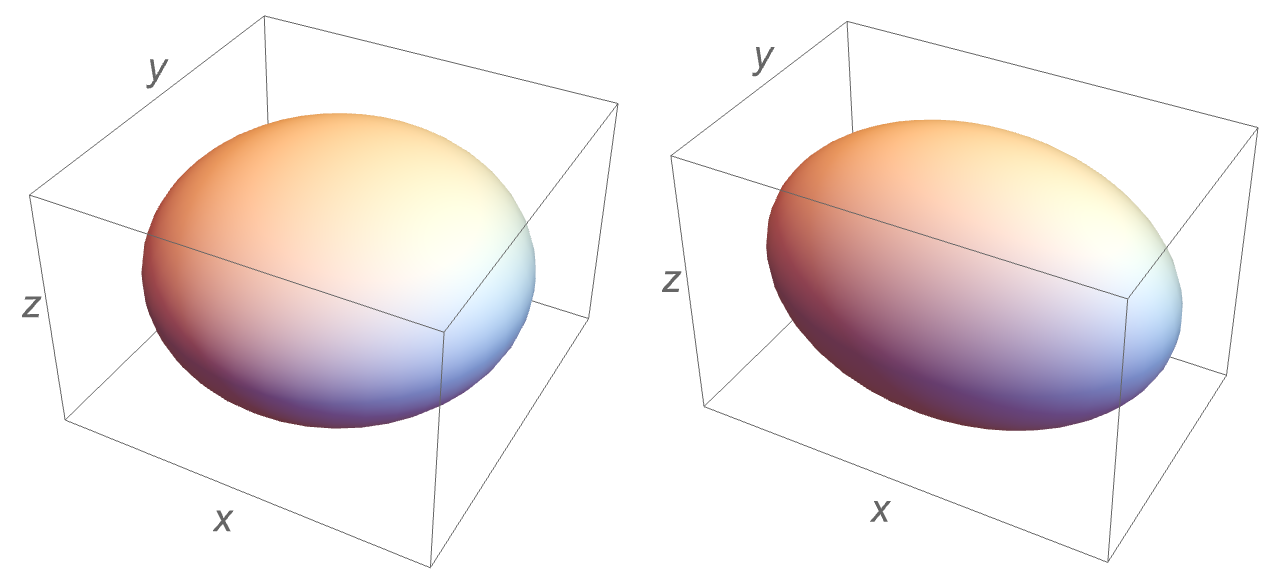}
\caption{\label{fig:ellipsoid} The surface of nonspherically closed domain walls in the cases with oblate (left) and prolate (right) ellipsoids. In this figure, the ratio of the semi-major axis (along $x$-axis) to the semi-minor axis (along $z$-axis) is 1.5.}
\end{figure}

We consider the static closed domain wall as the initial configuration for the scalar field. The simplest case is the spherical shell given by 
\begin{align} \label{eq:init_sph}
\phi(r) = v \tanh \left( \sqrt{\frac{\lambda}{2}} v (r - R_0) \right),\quad r = \sqrt{x^2 + y^2 + z^2},
\end{align}
with $R_0$ the initial shell radius introduced in Sec.~\ref{subsec:PBH}.
We also consider the nonspherical shape for the initial wall configuration.
The next-to-simplest case is the ellipsoid (spheroid), in which the initial scalar field configuration is given by
\begin{align} \label{eq:init_nonsph}
    \phi(x,y,z) = v \tanh\left[ \sqrt{\frac{\lambda}{2}} v \left(r - \frac{r}{\sqrt{(x/a)^2 + (y/b)^2 + (z/c)^2}} \right) \right],
\end{align}
where $a,b,c$ denote the semiaxes of ellipsoid along the $x,y,z$ axes respectively, that parametrize the nonsphericity.
Specifically, we consider the two cases: the oblate (pancake-like) ellipsoid with $a=b>c$, and the prolate (rugby-ball-like) ellipsoid with $a>b=c$, as illustrated in Fig.~\ref{fig:ellipsoid}.
The initial value for the conjugate momentum, $\Pi$, is set to zero everywhere in both spherical and nonspherical cases.

After setting the initial values for the scalar field, we set the initial values for the metric variables that satisfy both the Hamiltonian and the momentum constraints. 
Here, we set the initial value of the extrinsic curvature to zero. Then, the momentum constraint is trivially satisfied. 
Moreover, we assume the conformal flatness, that is, $\tilde{\gamma}_{ij} = \delta_{ij}$ on the initial spatial hypersurface $\Sigma_0$. 
Then, the remaining degree of freedom is only the conformal factor, $\chi$. To satisfy the Hamiltonian constraint, we solve the following nonlinear elliptic differential equation for the conformal factor,
\begin{align}
    \Delta \psi + 2\pi \psi^5 \rho = 0, \quad \psi=\chi^{-1/4},
\end{align}
where $\Delta$ represents the Laplacian in a flat space.
We solve this equation using the multigrid method \cite{Bentivegna:2013xna,tomida2023athena++} with a fixed boundary condition, $\psi_{\rm bd} = 1$ corresponding to the asymptotic flatness. After sufficient multigrid cycles, we get the initial configuration that satisfies the above equation with an error less than $10^{-10}$ at each site.

\subsection{Numerical implementation}

In our simulations, the spatial derivative is discretized by the central difference with the fourth-order accuracy \cite{Zlochower:2005bj}, except for the advection term with the shift vector for which we adopt the lopsided upwind difference with the fourth-order accuracy.
The field variables are updated using the fourth-order Runge-Kutta method \cite{press2007numerical}. The Kreiss-Oliger dissipation term \cite{KreissOliger1972} is added for the evolution of each dynamical field.

Simulating the black hole formation requires high resolution in a small limited region, but the region far from the black hole does not need such high resolution. Then, we employ the adaptive mesh refinement (AMR) technique, in which mesh is refined adaptively where higher resolution is requested.
Namely, the region where the conformal factor and the lapse take small values indicates the existence of a black hole, and then the mesh should be refined there.
Specifically, we apply the octree-based AMR, adopted in e.g. \texttt{Athena++} \cite{stone2020athena++} (and its variants \texttt{GR-Athena++} \cite{Daszuta:2021ecf} and \texttt{AthenaK} \cite{Zhu:2024utz})\footnote{
This is an alternative way from the block-structured (patch-based) AMR adopted in \texttt{GRChombo} \cite{Clough:2015sqa,Radia:2021smk}.
}.
The mesh is refined by separating a domain (parent box) into octants (eight boxes in the 3-dimensional space) where the refinement criterion is met. Each octant has the same number of grid points as the parent box. 
Conversely, the mesh is derefined in the domain where the criterion is no longer satisfied. See \cite{stone2020athena++} for details.
In our setup, the physical quantities are prolonged (restricted) by the fifth(fourth)-order Lagrange interpolation when the mesh is refined (derefined) and we allow for the maximum 9-level refinements.
The number of grid points in each box with any refinement level is $N^3=16^3$. For instance, a box of any size in Fig.~\ref{fig:snapshot1} contains $16^3$ grid points.
All physical quantities are put at the cell center. 

The size of the whole simulation box is $L = 128 v^{-1}$. It should be much larger than the initial size of the closed wall because we assume the asymptotic flatness in the initial condition.
In the main evolution phase, we impose the periodic boundary condition\footnote{
In general, the periodic boundary condition causes the cosmic expansion \cite{Yoo:2018pda}. However, because the duration of our simulation is not long and the simulation box is large enough, the system is hardly affected by the cosmic expansion.
}.
It is valid as long as we terminate the simulation before the boundary effect contaminates the dynamical domain of our simulation.
Indeed, the duration of our simulation is shorter than the half-light-crossing time, and thus the boundary effects can be safely ignored.
Note that the discrepancy between the boundary conditions for initial values and the main evolution causes sizable constraint violations, but they are suppressed at later time without reaching the dynamical domain, thanks to the large boxsize and the CCZ4 formulation. See Appendix~\ref{sec:validity}.

\section{Numerical results} \label{sec:numerical}

\subsection{Spherical collapse}

\begin{figure}[tbp]
\centering 
\includegraphics[width=5cm,clip]{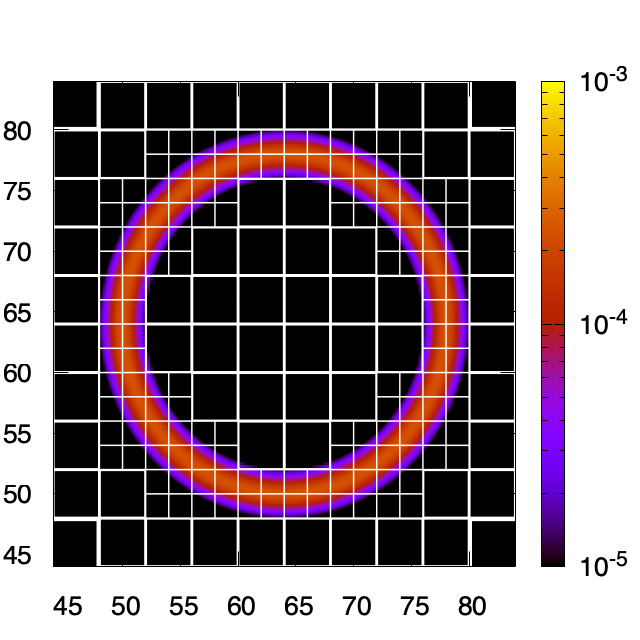}
\includegraphics[width=5cm,clip]{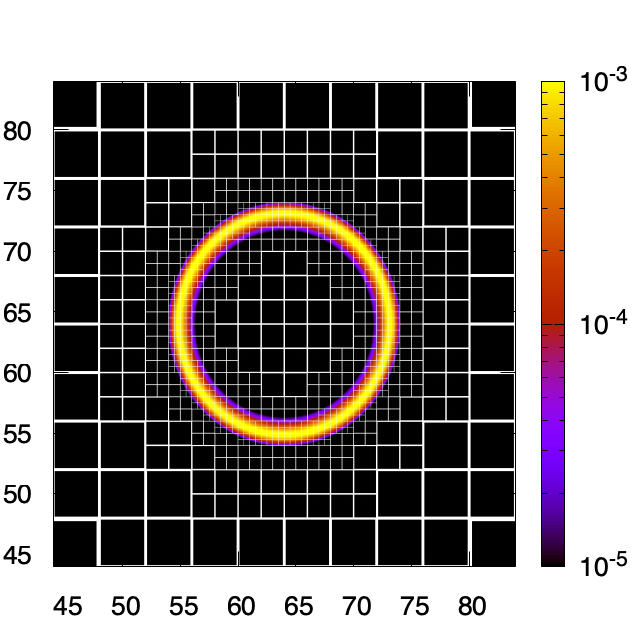}
\includegraphics[width=5cm,clip]{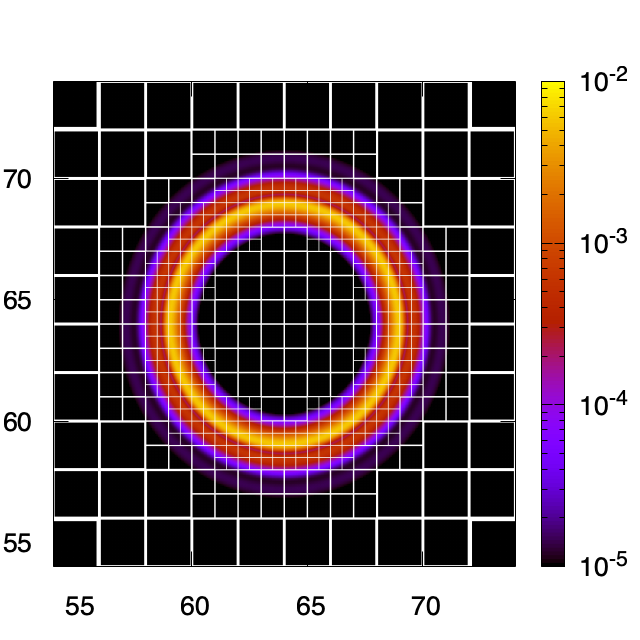}
\includegraphics[width=5cm,clip]{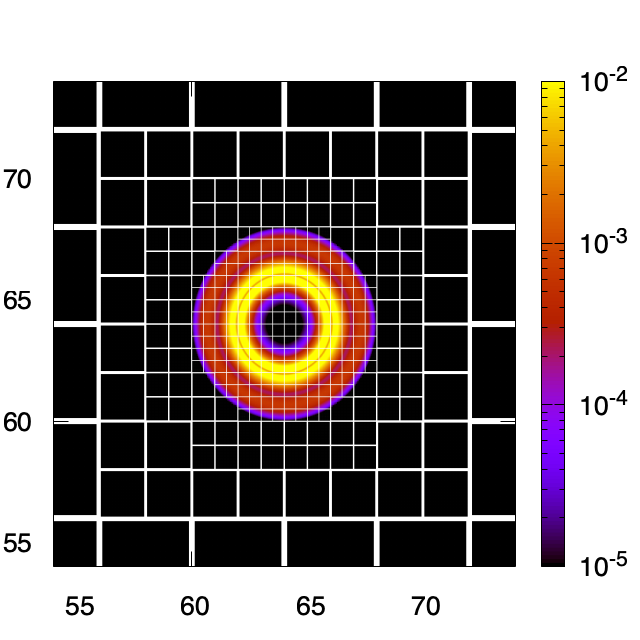}
\includegraphics[width=5cm,clip]{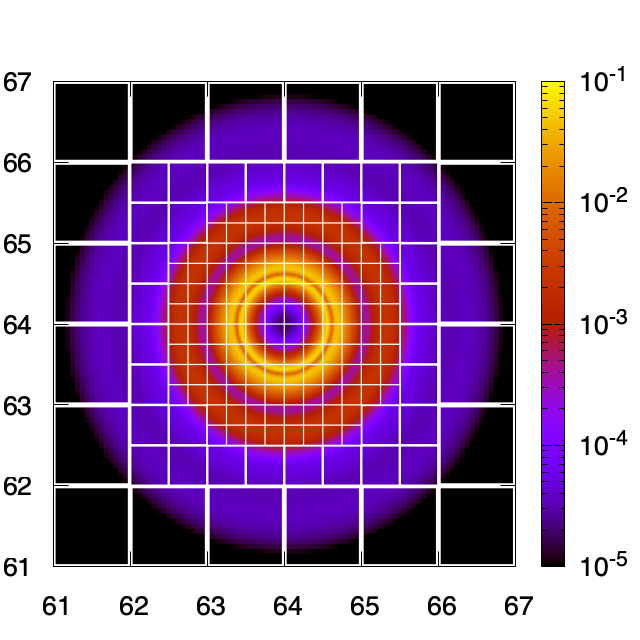}
\includegraphics[width=5cm,clip]{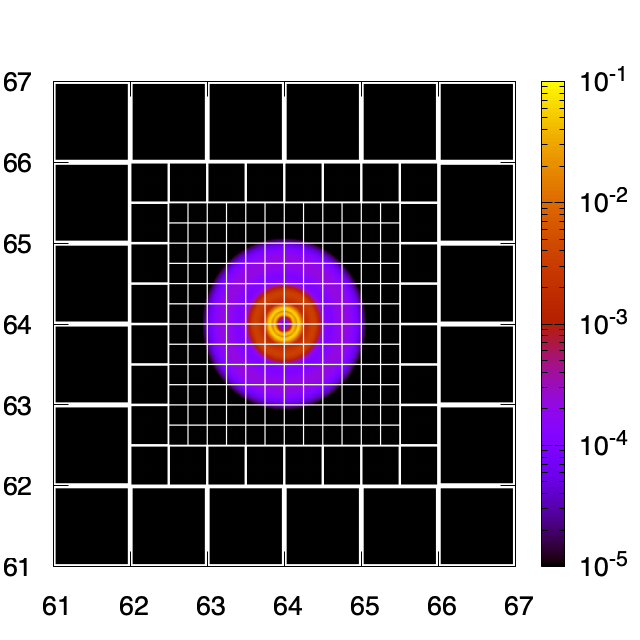}
\caption{\label{fig:snapshot1}
Time evolution of the 2-dimensional profile of the scalar field energy density.
Time evolves from upper left to upper right and then from lower left to lower right, corresponding to the coordinate time $vt = (0,10,15,20,25,30)$. Each axis and the color bar are respectively normalized by $v^{-1}$ and $v^{-4}$. Note that the scales of the axes and the color bar are different in each panel. We set $R_0 = 14v^{-1}$ and $v = 0.11M_{\rm pl}$. 
}
\end{figure}

Here, we show the results of our simulations with the spherical initial condition (\ref{eq:init_sph}).
We have taken various values for $v$ and $R_0$, and clarified whether the wall collapses into a black hole or not by evaluating the conformal factor and the lapse at the center of the simulation box. We also confirmed the formation of black holes using the apparent horizon finder with the multigrid elliptic solver \cite{Hui:2024ggb}.

\begin{figure}[tbp]
\centering 
\includegraphics[width=7.5cm,clip]{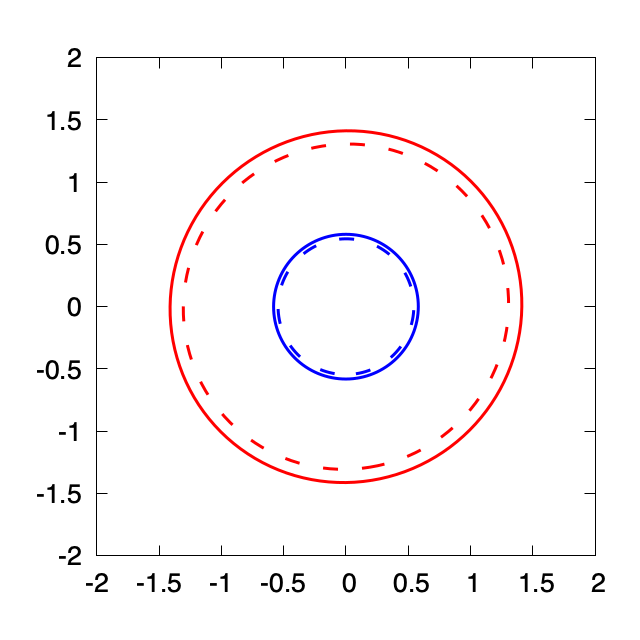}
\caption{\label{fig:AH} The 2-dimensional projection of the apparent horizon at the final time of the simulation $t=30v^{-1}$. 
Solid-red, solid-blue, dashed-red, dashed-blue correspond respectively to $(v/M_{\rm pl},\,vR_0) = (0.15,\,10),~(0.11,\,10),~(0.11,\,14),~(0.079,\,14)$.
}
\end{figure}

Fig.~\ref{fig:snapshot1} shows the time evolution of the 2-dimensional profile of the scalar field energy density. 
The wall thickness is clearly contracted as the shell radius is getting smaller.
In addition, the upper right panel shows that the wall configuration is significantly deformed from the original shape, as the energy gradient is clearly asymmetric inside and outside the wall. 
We found the apparent horizon at $t=23v^{-1}$ and thus, the apparent horizon exists at the center in the lower middle and lower right panels.
Those panels also show that the scalar field is accreting onto a black hole, and at the final stage of the simulation (lower right panel), it is captured within the horizon whose radius is about 1 (see the dashed red curve in Fig. \ref{fig:AH}). 
See also Appendix~\ref{sec:bounce} for the case of the bounce (no PBH formation).

\begin{figure}[tbp]
\centering 
\includegraphics[width=6.5cm,clip]{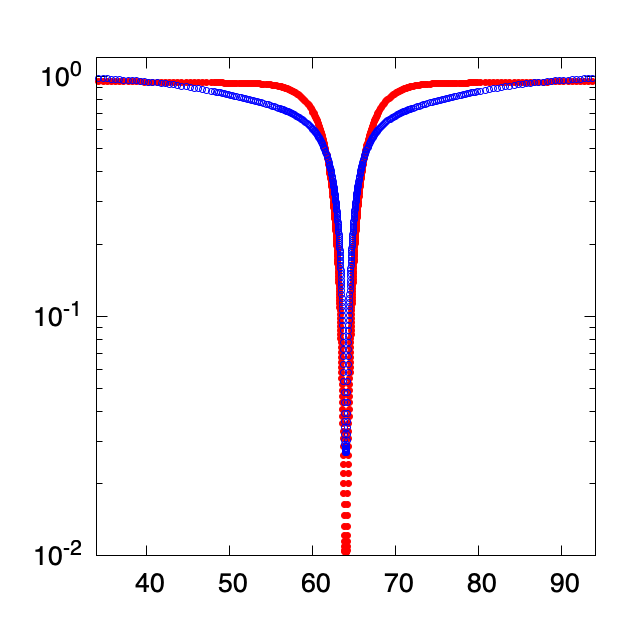}
\includegraphics[width=6.5cm,clip]{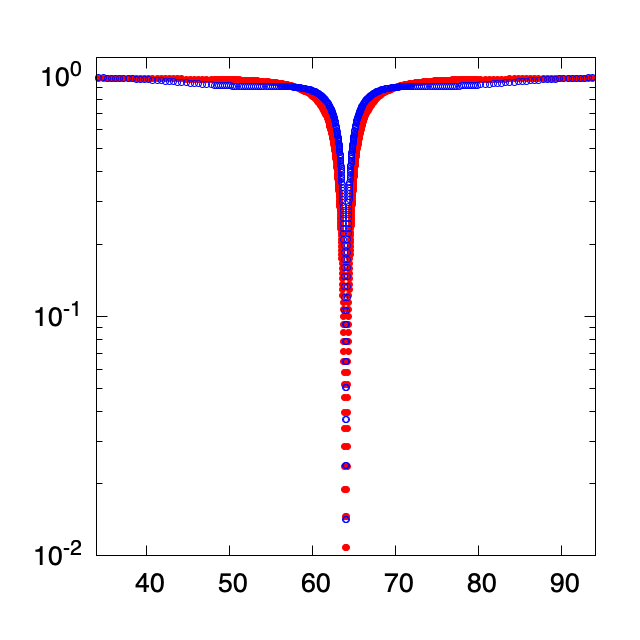}
\caption{\label{fig:chi} The profile of the conformal factor $\chi$ (red filled circle) and the lapse $\alpha$ (blue open circle) at the final time of the simulation $t=30v^{-1}$. We set $R_0=14v^{-1}$ and $v=0.11M_{\rm pl}$ ($0.079M_{\rm pl}$) the left (right) panel.}
\end{figure}

\begin{figure}[tbp]
\centering 
\includegraphics[width=7.5cm,clip]{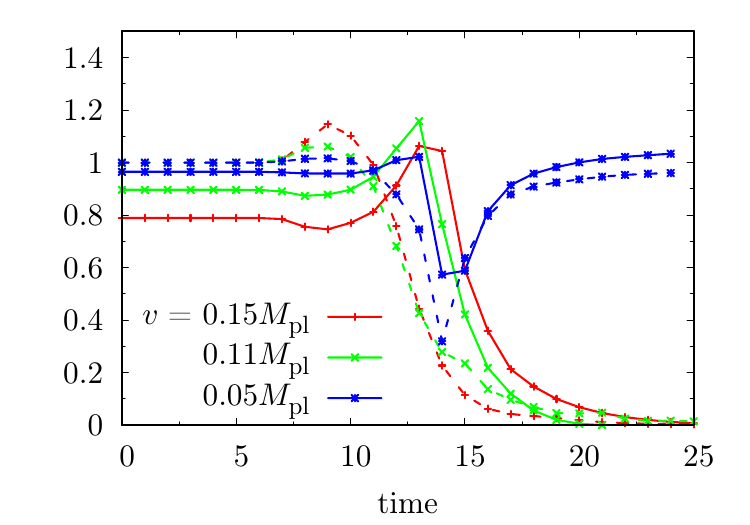}
\includegraphics[width=7.5cm,clip]{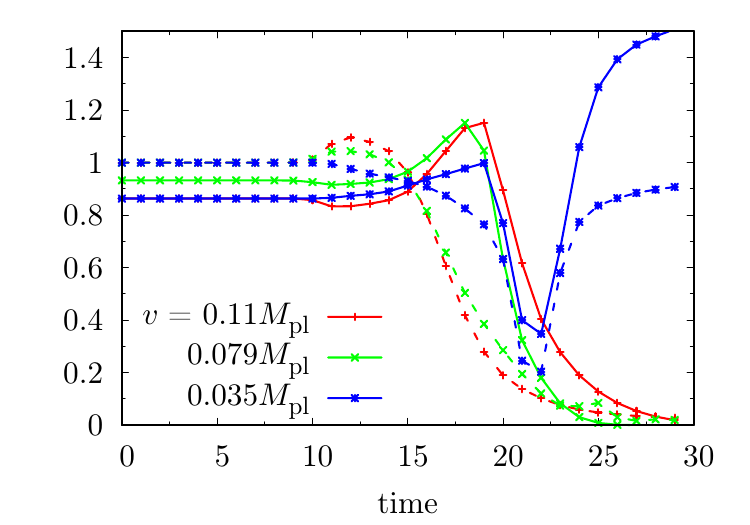}
\caption{\label{fig:central} The time evolution of the conformal factor $\chi$ (solid) and the lapse $\alpha$ (dashed) evaluated at the center of the simulation box. The initial shell radius is $R_0 = 10v^{-1}$ ($14v^{-1}$) in the left (right) panel.}
\end{figure}

Fig.~\ref{fig:chi} shows the profiles of the conformal factor, $\chi$, and the lapse, $\alpha$, at the final time of the simulation, $t=30v^{-1}$.
The figure demonstrates that both quantities drop steeply toward the center, indicating the existence of a black hole.
The time evolutions of those values at the center of the simulation box are shown in Fig.~\ref{fig:central}.
Those quantities fall suddenly at some time and approach zero in the cases with larger $v$ (corresponding to the red and green lines) but bounce otherwise (blue lines), showing there is a certain threshold for the PBH formation.
Fig.~\ref{fig:AH} shows the 2-dimensional projection of the apparent horizon at the final time of the simulation. 
Note that each set of solid and dashed curves has the same initial energy, $E_{\rm DW} = 4 \pi R_0^2 \sigma_{\rm DW}$. The figure indicates that the larger initial shell radius results in slightly smaller horizon size and thus a lighter black hole. This might be because the wall is more accelerated for the larger initial shell radius, and thus the plane wall profile is deformed more significantly.
Then, the outer part of the wall is partially detached before the black hole formation, as shown in the lower left panel in Fig.~\ref{fig:snapshot1}, which results in the energy loss and the smaller black hole mass.

\begin{figure}[tbp]
\centering 
\includegraphics[width=12cm,clip]{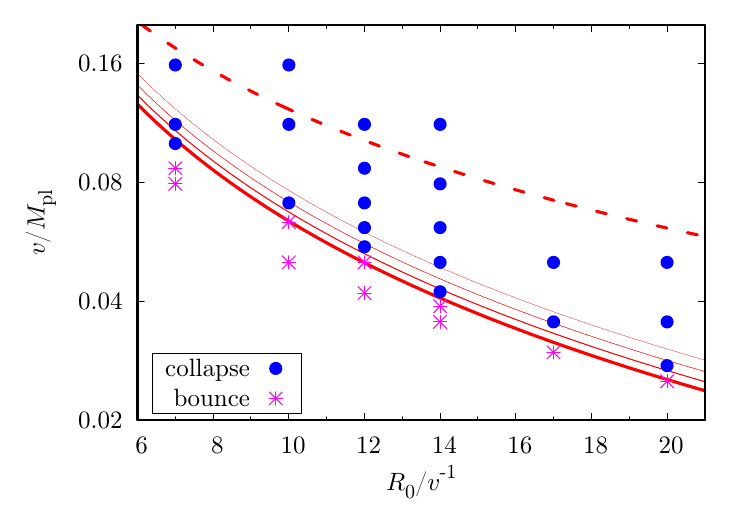}
\caption{\label{fig:BHcr} A viable parameter space for the PBH formation.
The blue and magenta points show the case with the collapse and the bounce.
The thick red solid and dashed curves represent, respectively, the lower bounds of (\ref{eq:PBHcriterion}) and (\ref{eq:PBHcriterion_naive}) with $f=1$. The thin solid curves correspond to the lower bound of (\ref{eq:PBHcriterion}) with $f=0.9,\,0.8,\,0.7$ from bottom to top.}
\end{figure}

Fig.~\ref{fig:BHcr} is a summary of our numerical results, showing a viable parameter space for the PBH formation. 
The blue and magenta points represent the cases with the collapse (PBH formation) and the bounce (no PBH) respectively.
The thick red solid and dashed curves correspond to the theoretical lower bounds for the PBH formation given by (\ref{eq:PBHcriterion}) and (\ref{eq:PBHcriterion_naive}) with $f=1$. The thin solid lines correspond to the lower bound of (\ref{eq:PBHcriterion}) with $f=0.9,\,0,8,\,0.7$ from bottom to top.
Our numerical results support the analytic consideration for the PBH formation with the Lorentz contraction \cite{Dunsky:2024zdo}.
The border of the collapse and the bounce lies on the curve with $f=0.8$~-~1, indicating that the collapse fraction is more than 80\%, which may depend on the values of $v$ and $R_0$.

\begin{figure}[tbp]
\centering 
\includegraphics[width=5cm,clip]{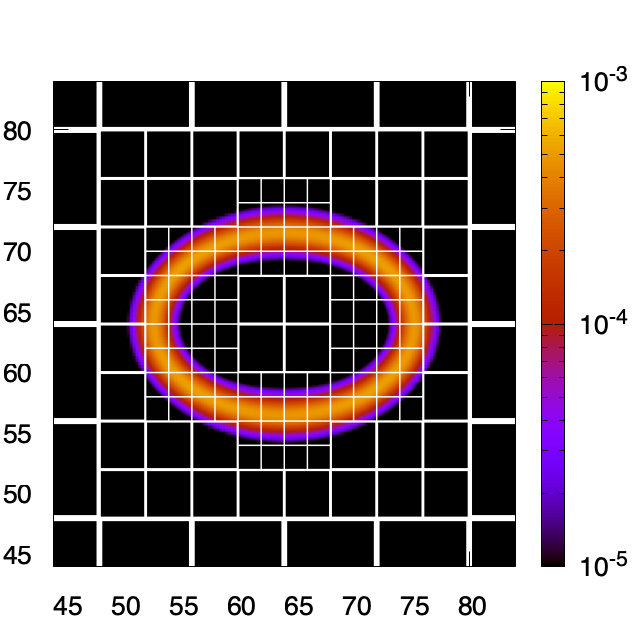}
\includegraphics[width=5cm,clip]{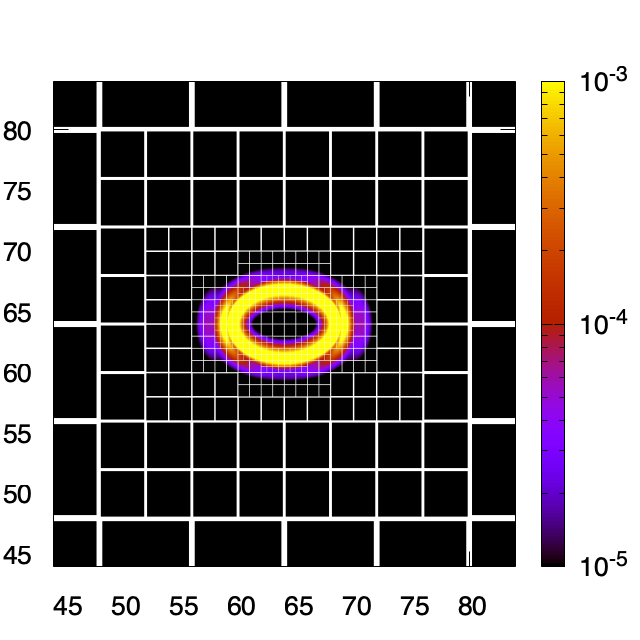}
\includegraphics[width=5cm,clip]{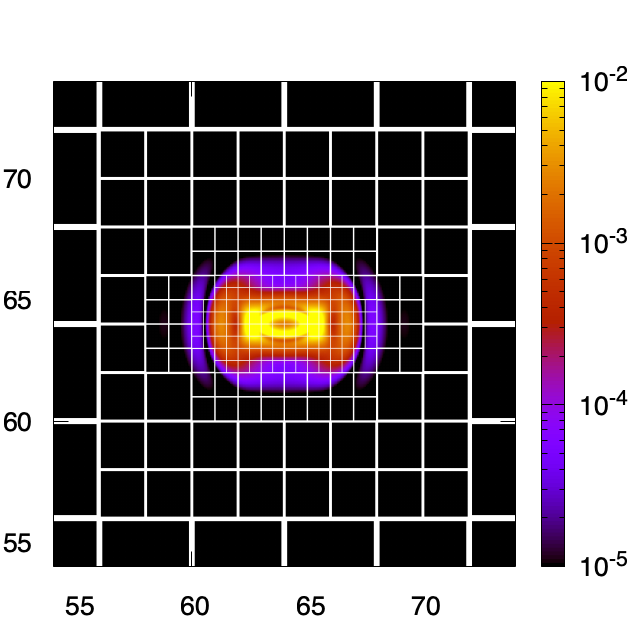}
\includegraphics[width=5cm,clip]{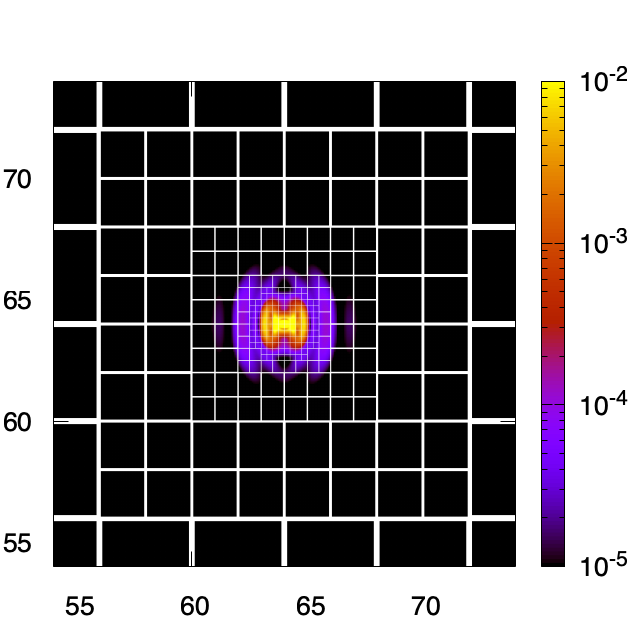}
\includegraphics[width=5cm,clip]{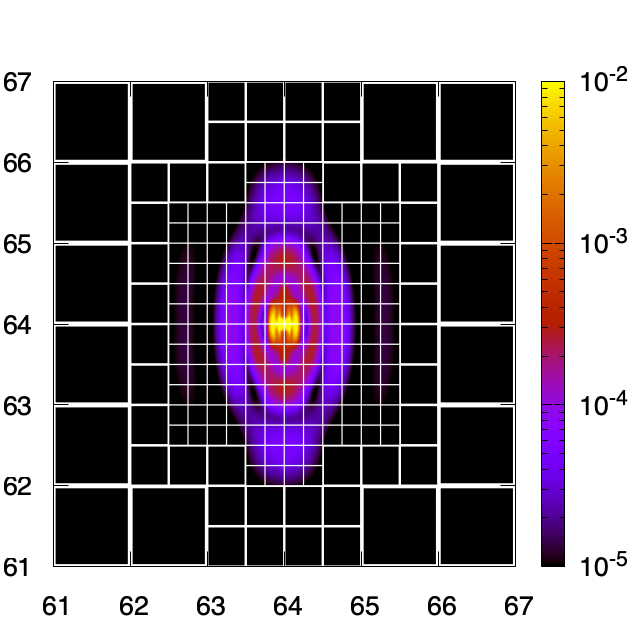}
\includegraphics[width=5cm,clip]{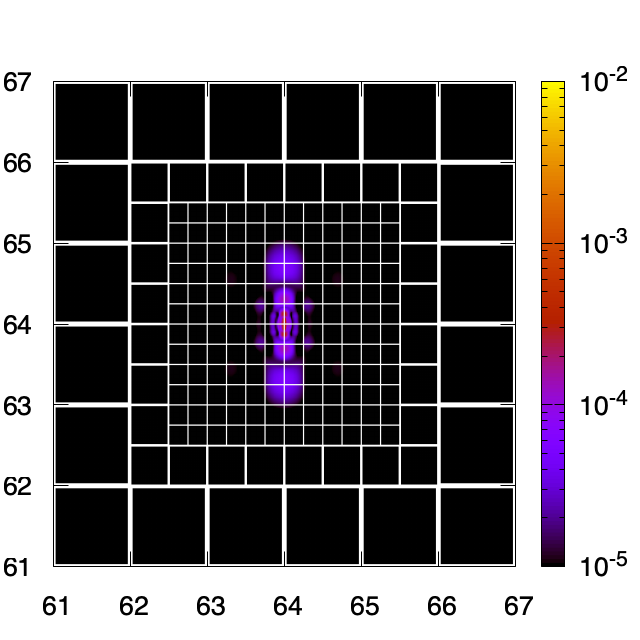}
\caption{\label{fig:snapshot_o} 
The same as Fig.~\ref{fig:snapshot1} but with the oblate ellipsoidal initial condition with $a/c=1.5$ and $v=0.16M_{\rm pl}$.
}
\end{figure}

\subsection{Non-spherical collapse}

Here, we show the numerical results in the case of the nonspherical collapse with the ellipsoidal initial condition (\ref{eq:init_nonsph}).
We specify the ratio of the semi-major to semi-minor axes, $a/c$, for the initial wall configuration and the lengths of these semi-axes are determined so that the surface area is equal to that of the sphere with $R_0=10v^{-1}$.
We examined five cases with $a/c = (1.1,\,1.2,\,1.3,\,1.4,\,1.5)$ for both oblate and prolate ellipsoids. 

\begin{figure}[tbp]
\centering 
\includegraphics[width=7.5cm,clip]{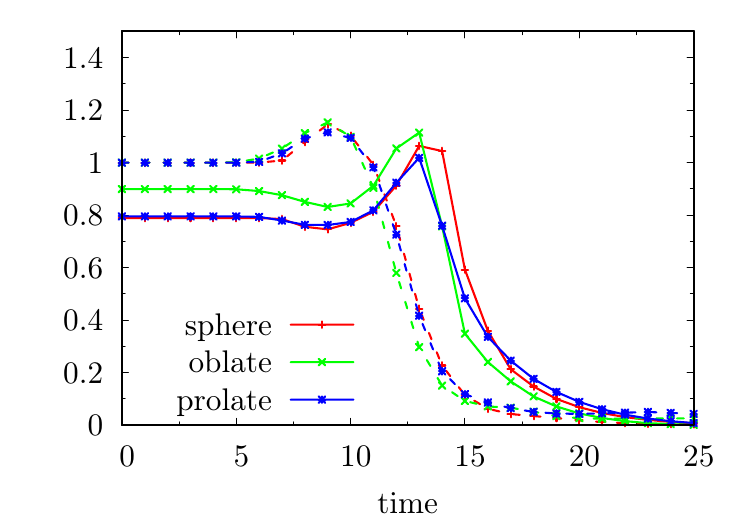}
\includegraphics[width=7.5cm,clip]{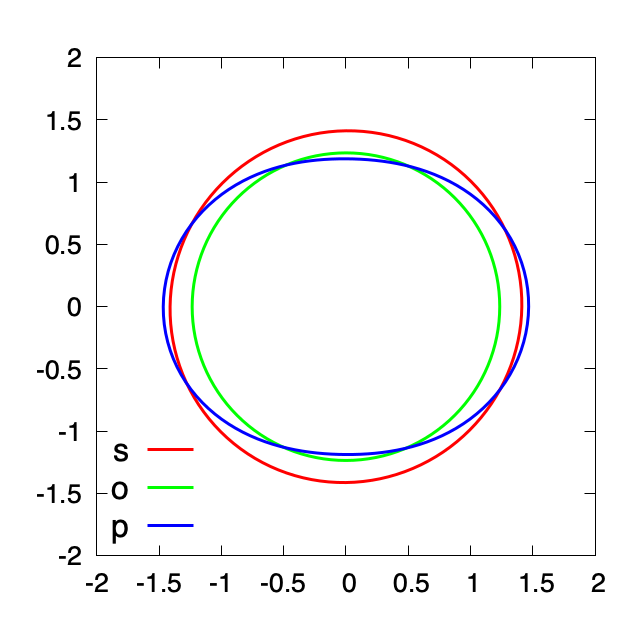}
\caption{\label{fig:nonSph} Left: Time evolution of the central values of the conformal factor $\chi$ and the lapse $\alpha$. Right: The 2-dimensional projection of the apparent horizon at the final time of the simulation $t=30 v^{-1}$.
In both panels, we have taken $a/c=1.5$ and the red, green and blue curves correspond respectively to the spherical, oblate ellipsoidal and prolate ellipsoidal cases.}
\end{figure}

We found the formation of black holes in all of the above cases. 
Fig.~\ref{fig:snapshot_o} exhibits the process of gravitational collapse with the oblate initial condition with $a/c=1.5$. (See Appendix~\ref{sec:bounce} for the prolate case.)
The dynamics shows a significant deviation from the spherical case, but we confirmed black hole formation by detecting the apparent horizon at $t=17v^{-1}$.
The left panel of Fig.~\ref{fig:nonSph} shows the evolutions of $\chi$ and $\alpha$ with $a/c = 1.5$ evaluated at the center of the simulation box. 
In both the oblate and prolate cases, the behavior is similar to the spherical case. The right panel of Fig.~\ref{fig:nonSph} shows the 2-dimensional projection (the $x$-$z$ plane in Fig.~\ref{fig:ellipsoid}) of the apparent horizon at the final time $t=30v^{-1}$. 
The apparent horizon in the oblate ellipsoidal case is clearly smaller than that in the spherical case. This may be partly because the system loses energy through the emission of gravitational waves as a significant deformation from the sphericity allows it.
In the prolate ellipsoidal case, the apparent horizon still highly deviates from the sphere. More detailed analysis is required to clarify the dependence on the nonsphericity parameters (such as the ellipticity and the prolateness) for the PBH formation.

\section{Discussion} \label{sec:discussion}

In this paper, we have investigated the PBH formation from the collapse of closed domain walls using numerical simulations with full general relativity. In particular, we have clarified the criteria for gravitational collapse with various values of the tension and initial radius (surface area) of the closed wall.
In particular, in the case of spherical collapse, we have shown that the viable parameter space for the PBH formation is consistent with the naive criterion based on a simple comparison between the Schwarzschild radius and the Lorentz-contracted wall thickness.
This supports the previous study based on semianalytic considerations and flat-space lattice simulations \cite{Dunsky:2024zdo}.
It also turned out that more than 80\% of the initial energy stored in the wall falls into a black hole.

We have also studied the non-spherical collapse by considering the ellipsoidal initial configurations for closed domain walls.
Our numerical results indicate that even when the ratio of the semi-major to semi-minor axes is as large as 1.5, black holes can form in both the oblate and prolate ellipsoidal cases.
However, the apparent horizon is smaller than that in the spherical case, implying the significant energy loss. This may be partly due to the emission of gravitational waves, since the dynamics shows strong nonsphericity, allowing for quadrupole radiation.
More detailed study is required for the PBH formation criterion and the collapse fraction in the nonspherical case, which is left for future work.

In this paper, we have not taken into account the background cosmic fluid that drives the cosmic expansion. In the cosmological context, our situation corresponds to the universe dominated by the scalar field constituting domain walls. In reality, however, the background cosmic fluid cannot be negligible in most cases and should therefore be included in the simulation, as is done in \cite{Deng:2016vzb}.
Then, we will be able to estimate quantitatively the PBH abundance in our universe from the initial abundance of closed walls \cite{Dunsky:2024zdo}.
Typically, the PBH formation requires superhorizon-scale closed walls. The probability of finding such walls is exponentially suppressed in ordinary cases. However, if the formation of domain walls is originated from the primordial inflationary (scale-invariant) fluctuations, such superhorizon-scale closed walls can be easily found \cite{Gonzalez:2022mcx,Kitajima:2023kzu}. 
In this case, the PBH might be more abundant or even overproduced.

In the model considered so far, the potential has two exactly degenerate minima. If we introduce a potential bias, the inward velocity of the wall can increase more efficiently due to an additional pressure exerted on the wall. 
Moreover, the false vacuum energy inside the wall increases the total energy of the system and thus the Schwarzschild radius and the mass of PBH.
Thus, the dynamics of the collapse and the PBH formation criterion can be significantly modified. The PBH formation in such a situation is briefly discussed in \cite{Ferreira:2024eru}, but we need detailed numerical analysis to calculate the PBH abundance more quantitatively.

Furthermore, if we relax the axial symmetry in the initial wall configuration and allow for more general triaxial ellipsoid (i.e. $a \neq b \neq c$), the system can possess a nonzero angular momentum, and the resultant black hole can be rotating \cite{Dunsky:2024zdo}.
It can be a characteristic signature because the spin of PBHs from primordial density fluctuations is typically suppressed \cite{Chiba:2017rvs,Harada:2017fjm,Mirbabayi:2019uph,DeLuca:2019buf,Harada:2020pzb,Saito:2023fpt,Saito:2024hlj,Ye:2025wif}.
The precise calculation of the PBH spin parameter necessitates a full numerical relativistic approach, which is also a future direction of our study.

\acknowledgments

We thank Sho Fujibayashi, Tomohiro Harada and Chul-Moon Yoo for helpful comments.
This work used computational resources of supercomputer AOBA at Cyberscience Center, Tohoku University, through JHPCN Joint Research Project (Project ID: jh250066).

\appendix

\section{CCZ4 formulation} \label{sec:ccz4}

In this paper, we adopt the CCZ4 formulation \cite{Alic:2011gg} for the evolution of the metric quantities. It is a variant form of the Z4 formulation \cite{Bona:2003fj} based on the BSSNOK formulation \cite{Nakamura:1987zz,Shibata:1995we,Baumgarte:1998te}, allowing constraint violating modes to propagate with damping. This formulation is based on the modification of the Einstein equation as follows
\begin{align}
^{(4)}R_{\mu\nu} + \nabla_\mu Z_\nu + \nabla_\nu Z_\mu = 8 \pi G \left(T_{\mu\nu} - \frac{1}{2} T g_{\mu\nu} \right) + \kappa_1 [ n_\mu Z_\nu + n_\nu Z_\mu - (1+\kappa_2)g_{\mu\nu} n_\alpha Z^\alpha],
\end{align}
where $^{(4)}R_{\mu\nu}$ is the 4-dimensional Ricci tensor, $Z_\mu$ is an additional dynamical field, $T$ is the trace of the energy momentum tensor, and $\kappa_1$ and $\kappa_2$ are numerical coefficients characterizing the damping of constraint violations. The original Einstein equation is restored for $Z_\mu = 0$, which is thus regarded as a constraint.
Let us rewrite the traceless part of the extrinsic curvature,
\begin{align}
A_{ij} = \chi^{-1} \tilde{A}_{ij},
\end{align}
and define the following quantities,
\begin{align}
\Theta = - n_\mu Z^\mu,\quad  \hat\Gamma^i = \tilde\Gamma^i + 2 \tilde\gamma^{ij} Z_j,
\end{align}
where $\tilde\Gamma^i = \tilde\gamma^{ij} \tilde\Gamma^i_{jk}$ with $\tilde\Gamma^i_{jk}$ the Christoffel symbol associated with the conformal metric $\tilde\gamma_{ij}$.
Then, the system of evolution equations is given as follows
\begin{align}
\partial_t \chi &= \frac{2}{3} \alpha \chi K -\frac{2}{3} \chi \partial_k \beta^k + \beta^k \partial_k \chi \\
\partial_t \tilde\gamma_{ij} & = -2\alpha \tilde{A}_{ij} + 2\tilde\gamma_{k(i} \partial_{j)} \beta^k - \frac{2}{3} \tilde\gamma_{ij} \partial_k \beta^k + \beta^k \partial_k \tilde\gamma_{ij},
\end{align}
\begin{align}
\begin{split}
\partial_t \tilde{A}_{ij} &= \chi \left[ -D_i D_j \alpha + \alpha (R_{ij} + D_i Z_j + D_j Z_i - 8 \pi G S_{ij}) \right]^{\rm TF} \\
& \quad + \alpha \left[ \tilde{A}_{ij} (K - 2 \Theta) - 2 \tilde{A}_{il} \tilde{A}^l_j \right] + 2 \tilde{A}_{k(i} \partial_{j)} \beta^k - \frac{2}{3} \tilde{A}_{ij} \partial_k \beta^k + \beta^k \partial_k \tilde{A}_{ij},
\end{split} \label{eq:dt_Aij_ccz4} \\[2mm]
\begin{split}
\partial_t K &= - D_i D^i \alpha + \alpha (R + 2 D_i Z^i + K^2 - 2 \Theta K) + \beta^i \partial_i K \\
& \quad- 3 \alpha \kappa_1 (1 + \kappa_2) \Theta + 4 \pi G \alpha (S - 3 \rho),
\end{split} \label{eq:dt_K_ccz4}
\end{align}
\begin{align}
\begin{split}
\partial_t \Theta &= \frac{1}{2} \alpha \left( R + 2 D_i Z^i - \tilde{A}_{ij} \tilde{A}^{ij} + \frac{2}{3} K^2 - 2 \Theta K \right) - Z^i \partial_i \alpha + \beta^k \partial_k \Theta  \\
& \quad - \alpha \kappa_1 (2 + \kappa_2) \Theta - 8 \pi G \alpha \rho
\end{split} \label{eq:dt_Theta_ccz4} \\[2mm]
\begin{split}
\partial_t \hat\Gamma^i &= - 2 \tilde{A}^{ij} \partial_j \alpha + 2 \alpha \left( \tilde\Gamma^i_{jk} \tilde{A}^{jk}  - \frac{3}{2\chi}\tilde{A}^{ij} \partial_j \chi -\frac{2}{3} \tilde\gamma^{ij} \partial_j K \right) \\
&\quad + 2 \tilde\gamma^{ij} \left( \alpha \partial_j \Theta - \Theta \partial_j \alpha - \frac{2}{3} \alpha K Z_j \right) \\
&\quad + \beta^k \partial_k \hat\Gamma^i  + \tilde\gamma^{jk} \partial_j \partial_k \beta^i + \frac{1}{3} \tilde\gamma^{ij} \partial_j \partial_k \beta^k + \frac{2}{3} \tilde\Gamma^i \partial_k \beta^k - \tilde\Gamma^k \partial_k \beta^i \\
&\quad + 2 \kappa_3 \left( \frac{2}{3} \tilde\gamma^{ij} Z_j \partial_k \beta^k - \tilde\gamma^{jk} Z_j \partial_k \beta^i \right) - 2\alpha \kappa_1 \tilde\gamma^{ij} Z_j - 16 \pi G \alpha \tilde\gamma^{ij} S_j,
\end{split}
\end{align}
together with the source terms,
\begin{align}
    \rho = n^\mu n^\nu T_{\mu\nu},\quad S_\mu = \gamma^\nu_\mu n^\lambda T_{\nu\lambda},\quad S_{\mu\nu} = \gamma^\alpha_\mu \gamma^\beta_\nu T_{\alpha\beta},\quad S = \gamma^{\mu\nu} S_{\mu\nu},
\end{align}
where $R_{ij}$ and $R$ are respectively the Ricci tensor and the Ricci scalar related to the spatial metric $\gamma_{ij}$, TF denotes the extraction of the trace-free part, and $\kappa_3$ is an additional coefficient to stabilize the numerical evolution.
In our simulations, we set (see also \cite{Radia:2021smk})
\begin{align}
    \kappa = 0.1/\alpha,\quad \kappa_2 = 0, \quad \kappa_3 = 1.
\end{align}
The Hamiltonian and momentum constraints are given by
\begin{align}
\mathcal{H} &= R + \frac{2}{3} K^2 - \tilde{A}_{ij} \tilde{A}^{ij} - 16 \pi G \rho = 0,\\
\mathcal{M}^i &= \chi \left( \tilde{D}_j \tilde{A}^{ij} - \frac{3}{2} \tilde{A}^{ij} \partial_j \ln \chi - \frac{2}{3} \tilde\gamma^{ij} \partial_j K \right) - 8 \pi G S^i = 0,
\end{align}
where $\tilde{D}_i$ is the covariant derivative associated with $\tilde\gamma_{ij}$.
In addition, we adopt the moving puncture gauge \cite{Campanelli:2005dd,Baker:2005vv} for the lapse and the shift vector.  
In this gauge, the encounter of singularities can be avoided by letting the lapse and the shift vector evolve in the following manner,
\begin{align}
    \partial_t \alpha &= - a_2 \alpha^{a_3}(K - 2 \Theta) + a_1 \beta^i \partial_i \alpha, \\
    \partial_t \beta^i &= b_2 B^i + b_1 \beta^j \partial_j \beta^i, \\
    \partial_t B^i &= c_2 \alpha^{c_3}(\partial_t \hat\Gamma^i - \beta^j \partial_j \hat\Gamma^i) - \eta B^i + c_1 \beta^j \partial_j B^i,
\end{align}
where $B^i$ is an auxiliary field and the numerical coefficients are chosen as follows
\begin{align}
    (a_1,a_2,a_3) = (1,2,1),\quad (b_1,b_2) = (0, 3/4),\quad (c_1,c_2,c_3) = (0,1,0),\quad \eta = 1.
\end{align}

\section{Validity check} \label{sec:validity}

\begin{figure}[tbp]
\centering 
\includegraphics[width=5cm,clip]{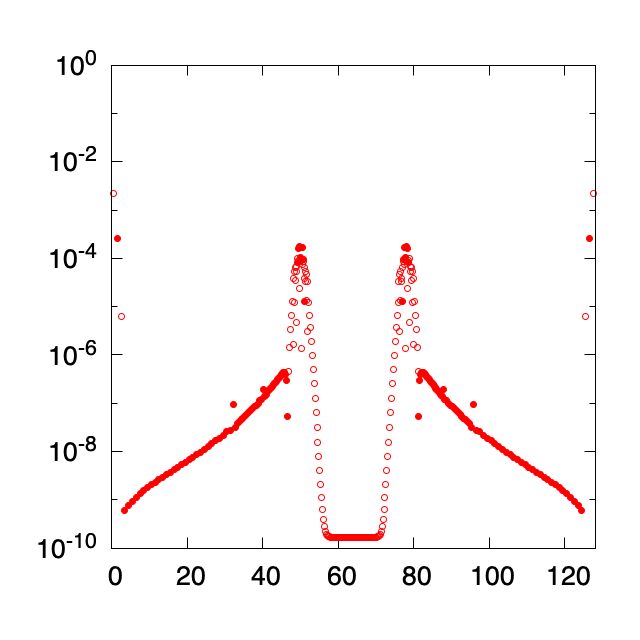}
\includegraphics[width=5cm,clip]{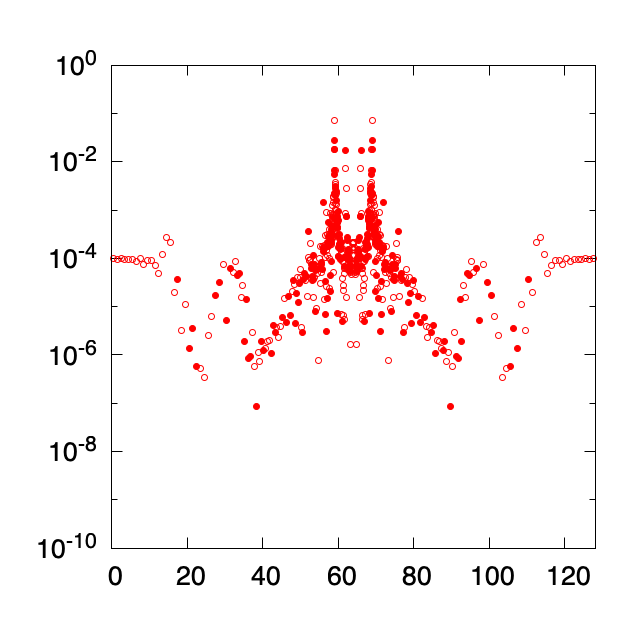}
\includegraphics[width=5cm,clip]{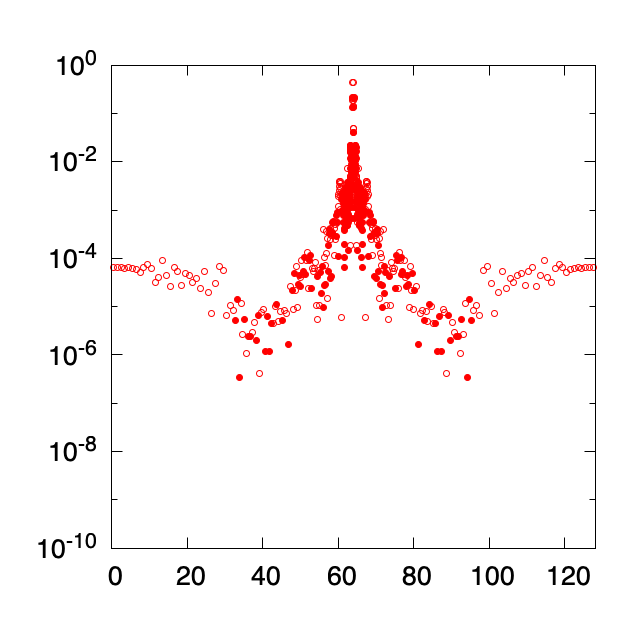}
\caption{\label{fig:constraint}The 1-dimensional profile of the local Hamiltonian constraint violation at three different time slices, $vt = 0,\,15,\,30$ from left to right, corresponding respectively to the top-left, top-right and bottom-right panels in Fig.~\ref{fig:snapshot1}. Open (closed) circles represent the positive (negative) value. We have taken $R_0 = 14v^{-1}$ and $v = 0.11M_{\rm pl}$.}
\end{figure}

To verify the validity of our simulations, we monitor violations of the Hamiltonian constraint, the momentum constraint, $\Theta = 0$, and $Z_i = 0$. The most stringent one is the Hamiltonian constraint, and Fig.~\ref{fig:constraint} shows the profile of the local Hamiltonian constraint violation on the axis containing the center of the simulation box. The left panel corresponds to the initial time and shows that the constraint is well satisfied except for the boundary. Two peaks correspond to the high energy region near the domain wall core. The middle panel corresponds to the time before the collapse. The constraint violation at the boundary is reduced due to the propagation and damping. 
The right panel corresponds to the final time of the simulation after the formation of a black hole. The violation is well suppressed outside the apparent horizon.

\begin{figure}[tbp]
\centering 
\includegraphics[width=5cm,clip]{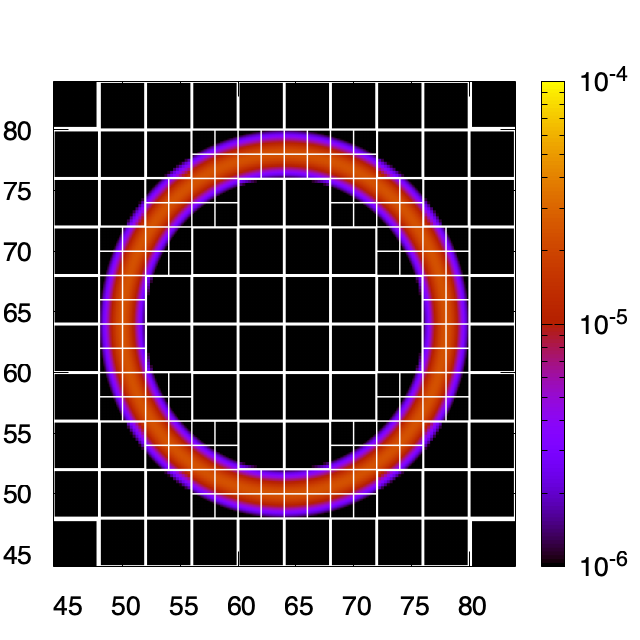}
\includegraphics[width=5cm,clip]{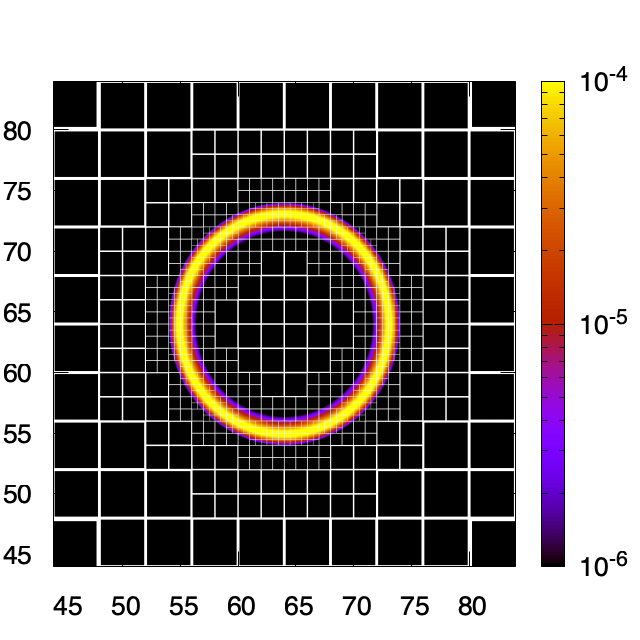}
\includegraphics[width=5cm,clip]{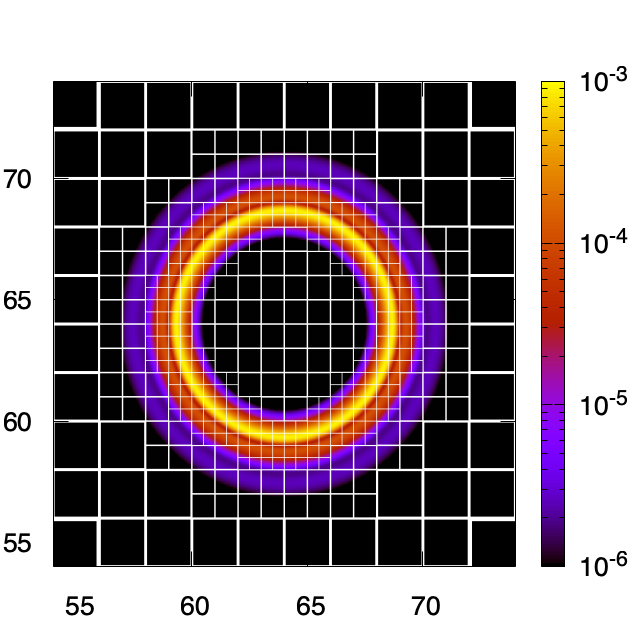}
\includegraphics[width=5cm,clip]{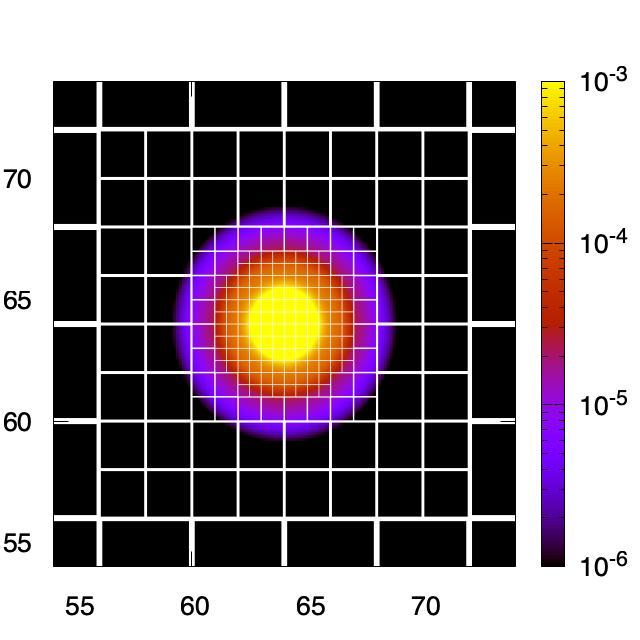}
\includegraphics[width=5cm,clip]{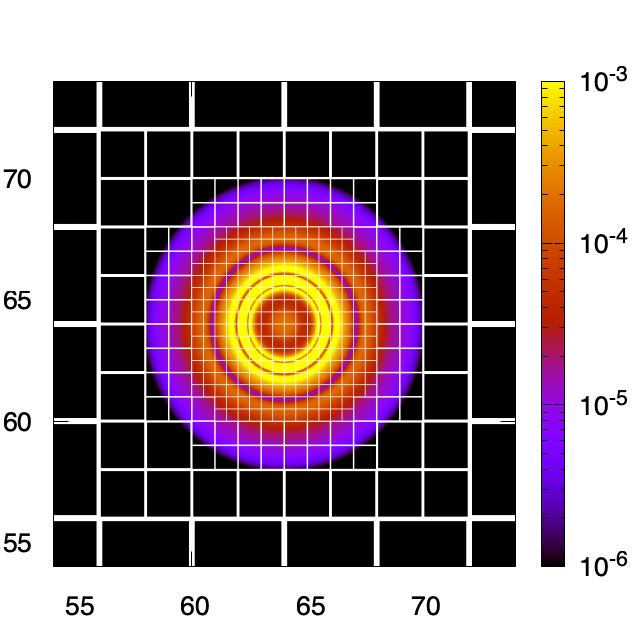}
\includegraphics[width=5cm,clip]{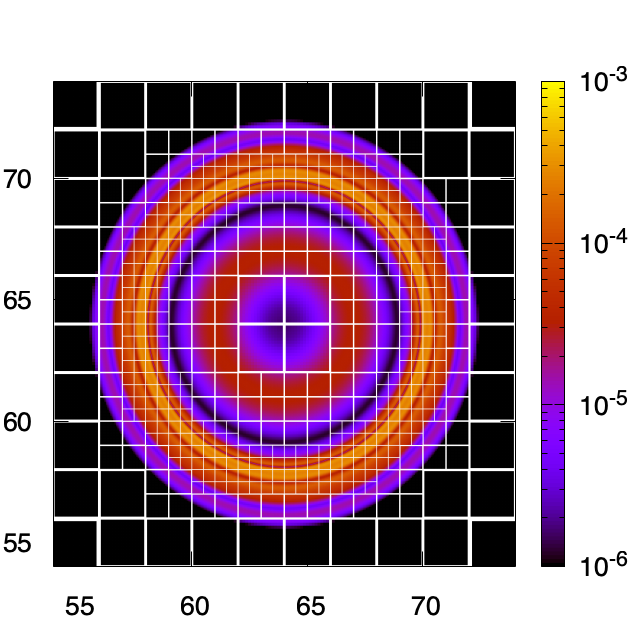}
\caption{\label{fig:snapshot2} 
The same as Fig.~\ref{fig:snapshot1} but $v = 0.035M_{\rm pl}$.
}
\end{figure}

\begin{figure}[tbp]
\centering 
\includegraphics[width=5cm,clip]{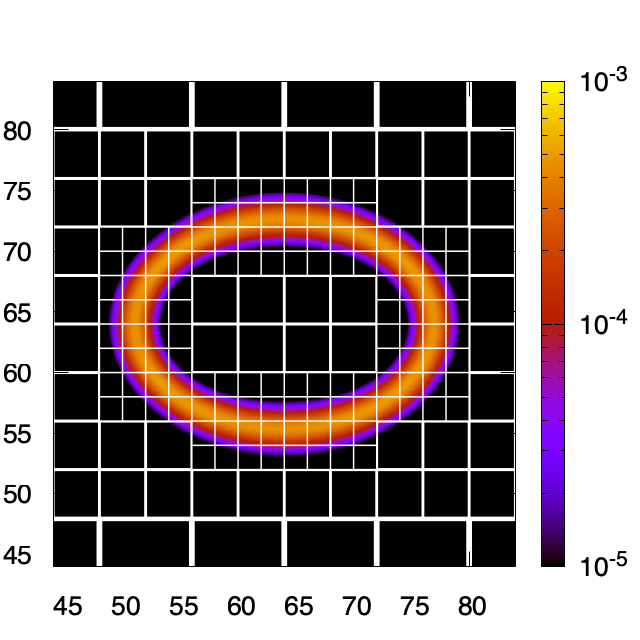}
\includegraphics[width=5cm,clip]{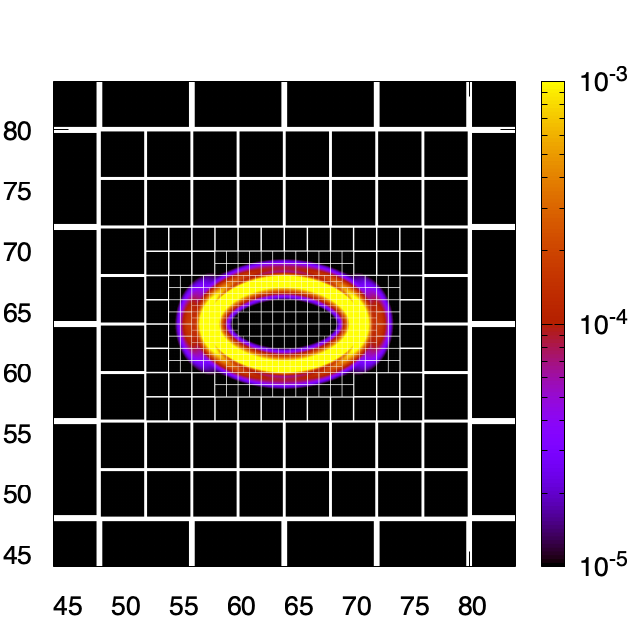}
\includegraphics[width=5cm,clip]{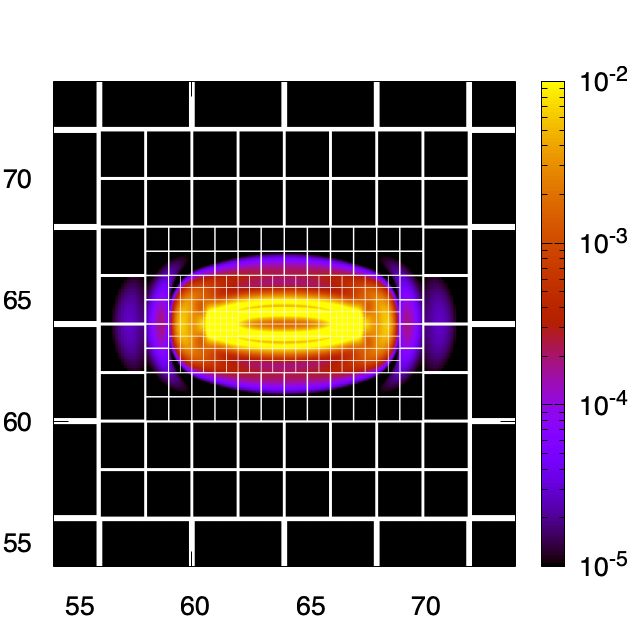}
\includegraphics[width=5cm,clip]{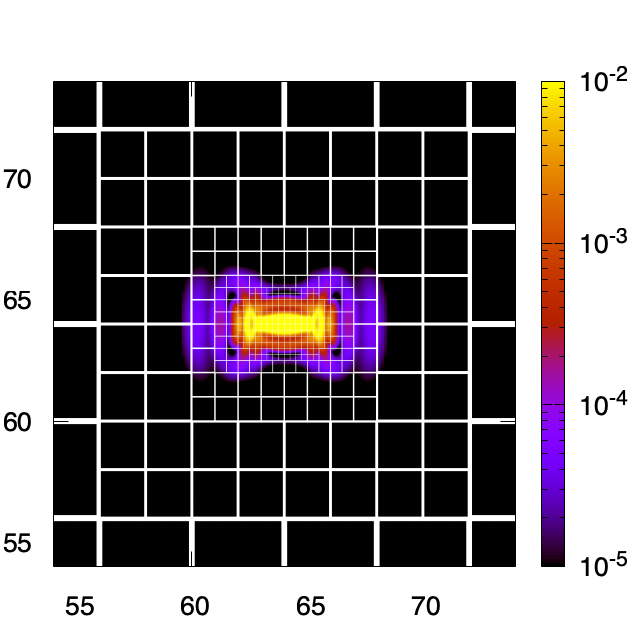}
\includegraphics[width=5cm,clip]{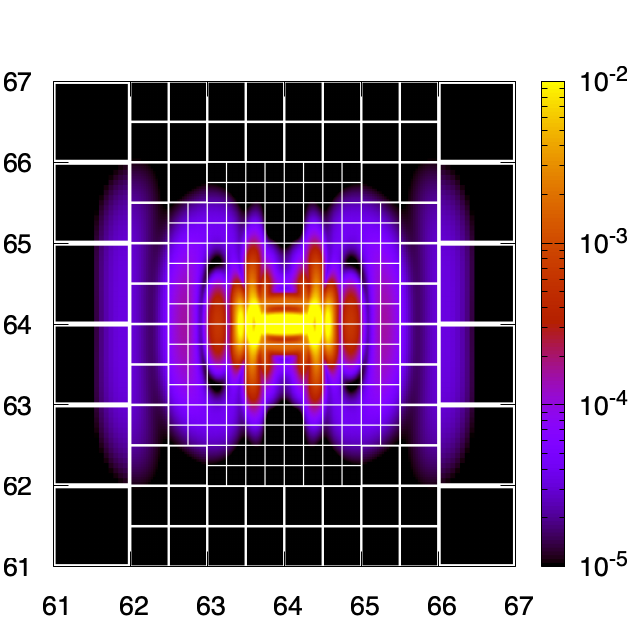}
\includegraphics[width=5cm,clip]{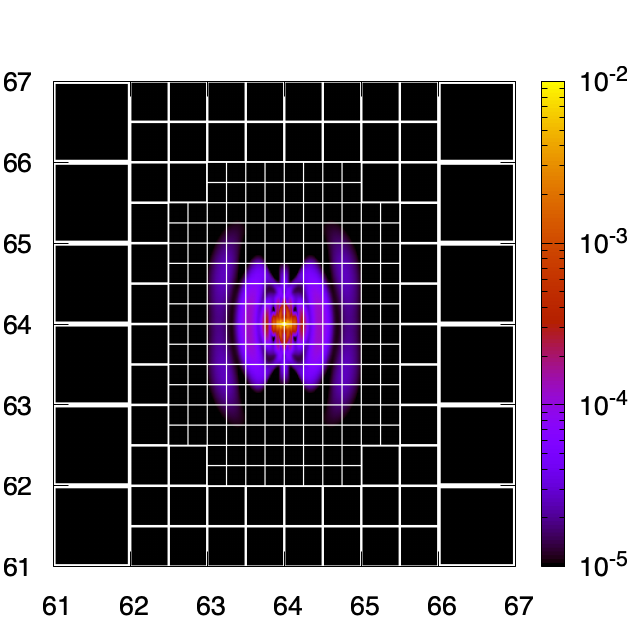}
\caption{\label{fig:snapshot_p} 
The same as Fig.~\ref{fig:snapshot_o} but with the prolate ellipsoidal initial condition.
}
\end{figure}

\section{Bounce and prolate ellipsoidal collapse} \label{sec:bounce}

Here we exhibit the 2-dimensional profiles of the scalar field energy density in the cases with the bounce and the prolate ellipsoidal collapse.
Fig.~\ref{fig:snapshot2} illustrates the time evolution in the bounce case. This situation corresponds to the blue line in the right panel of Fig.~\ref{fig:central}.
The wall falls toward the center and is maximally compressed, as shown in the bottom left panel. Then, it is bounced and the scalar field energy spreads out afterward, leaving no black hole.
The initial energy of the wall is not sufficient to generate a strong gravity to trap the imploded energy.

Fig.~\ref{fig:snapshot_p} exhibits the dynamics of the prolate ellipsoidal collapse. The semi-major and semi-minor axes are set as in the corresponding oblate case. 
First, the prolateness grows as the system evolves, and the wall develops into a spindle-like structure, as shown in the top right and bottom left panels\footnote{
Naked singularities may appear at the gravitational collapse of spindle-like objects \cite{Yoo:2016kzu}. Indeed, our simulation becomes unstable when we examine the case with higher prolateness, e.g. $a/c = 2$, probably due to the appearance of such singularities.
}.
Then, it collapses along the axial direction. The deviation from sphericity is more prominent than that in the oblate case. Finally, a black hole is left at the center, but the apparent horizon is still highly deformed from the sphere, as shown in the right panel in Fig.~\ref{fig:nonSph}.

\bibliographystyle{utphys}
\bibliography{ref}

\end{document}